\documentclass{DISproc}

\begin{document}
\title{Inclusive Jet Production Measured with ATLAS and Constraints on PDFs}

\author{{\slshape Bogdan Malaescu}, On behalf of the ATLAS Collaboration \\[1ex]
CERN, CH-1211, Geneva 23, Switzerland }



\maketitle

\begin{abstract}
Inclusive jet and dijet double-differential cross sections have been measured in proton-proton collisions at a centre-of-mass energy of 7~TeV using the ATLAS detector at the LHC. The cross sections were measured using jets clustered with the anti-$k_{t}$ algorithm. The measurements are performed in the jet rapidity range $\vert{\rm y}\vert<4.4$, covering jet transverse momenta from 20~GeV to 1.5~TeV and dijet invariant masses from 70~GeV to 5~TeV. The data are compared to expectations based on next-to-leading order QCD calculations corrected for non-perturbative effects, as well as to next-to-leading order Monte Carlo predictions. The data test the theory in a new kinematic regime, and provide sensitivity to parton distribution functions in a region where they are currently not well constrained.
\end{abstract}

\section{Introduction}

The inclusive jet and dijet cross sections are important tools for testing Quantum Chromodynamics~(QCD) and searching for physics beyond the Standard Model at the LHC.
The ATLAS Collaboration has published a first measurement of these cross sections at $\sqrt{s}=7~{\rm TeV}$, using an integrated luminosity of $17~{\rm nb}^{-1}$~\cite{Aad:2010ad}.
A second measurement, using the full 2010 data sample of $37.3~{\rm pb}^{-1}$~\cite{Aad:2011fc}, significantly extended the covered phase space: in the jet transverse momentum
${\rm p}_{\rm T}$~(from 60~GeV down to 20~GeV, and from 600~GeV up to 1.5~TeV), in rapidity~(from $\vert{\rm y}\vert<2.8$ to $\vert{\rm y}\vert<4.4$),
as well as in dijet mass~(from $1.8$ up to $4.8$~TeV).
A preliminary measurement of the dijet cross section using $4.7~{\rm fb}^{-1}$ in the full 2011 data sample~\cite{ATLASmoriond2012} shows an improved precision comparing to the previous
measurements.
These analyses probe next-to-leading order~(NLO) perturbative QCD and parton distribution functions~(PDFs) in a kinematic regime not explored before.

\section{Jet definition, reconstruction and calibration}

For the ATLAS inclusive jet and dijet cross section measurements, jets are defined using the anti-$k_{t}$ algorithm~\cite{Cacciari:2008gp}.
The measurements are performed for two different values of the distance parameter R~($0.4$ and $0.6$).

Jets are reconstructed at the electromagnetic~(EM) scale~\footnote{The EM-scale correctly reconstructs the energy of the electromagnetic showers deposited in the calorimeter.},
the inputs being three-dimensional topological clusters built from calorimeter cells.
The four-momentum of the uncalibrated, EM-scale jet is defined as the sum of the four-momenta of its constituent calorimeter energy clusters.
Additional energy due to pile-up interactions is subtracted by applying a correction depending on the number of reconstructed vertices in the event.
The energy of the jet is then corrected for instrumental effects like energy lost in the dead material or due to calorimeter non-compensation.
This jet energy scale~(JES) correction, as a function of the energy and pseudorapidity of the reconstructed jet, is derived using isolated jets in the Monte Carlo simulation~(MC)~\cite{Aad:2011he}.

The JES uncertainty is the dominant uncertainty for the inclusive jet and dijet measurements~\cite{Aad:2011fc}.
Comparing to the previous measurement~\cite{Aad:2010ad}, this uncertainty has been strongly reduced, due to an improved calibration of the calorimeter response at the EM-scale
using $Z \to ee$ \emph{in-situ} data, as well as using the single hadron energy measurement from \emph{in-situ} and test-beam data.
The improved precision is confirmed by independent \emph{in-situ} measurements in collision data, like the comparison of calorimeter jet energy to the sum of track ${\rm p}_{\rm T}$
associated to the jet, and transverse momentum balance in $\gamma+$jet, dijet and multijet events~\cite{Aad:2011he}.

In order to allow for a reliable treatment of the bin-to-bin correlations of the uncertainties, it is important to separate the different uncertainty components.
Therefore the JES uncertainty has beed split in several components, and the calorimeter component~(dominant in the central region) has been split in several sources.
All the uncertainty components~(sources) are treated as fully correlated in ${\rm p}_{\rm T}$ and rapidity, and independent between each other.

\section{Data correction to particle level}

The measured cross sections are corrected for the experimental effects and are hence obtained for the \emph{particle level} final state.
In MC particle jets are built from stable particles, including muons and neutrinos from decaying hadrons.

The inclusive jet and dijet measurements are corrected from detector to particle level using a matrix based unfolding~\cite{Aad:2011fc}.
A transfer matrix relating particle level and reconstructed quantities is built from MC, using a geometrical matching between particle level and reconstructed level jets.
The matching efficiency is taken into account in a three step unfolding procedure.
The first step applies the matching efficiency at the reconstructed level to the data spectrum, so that it can be directly compared with the spectrum of MC matched reconstructed jets.
The second step performs the actual unfolding, correcting for the transfer of jets~(events) between the bins.
Finally, the third step corrects for the matching efficiency at the particle level.

The bin-by-bin, SVD~\cite{Hocker:1995kb} and IDS~\cite{Malaescu:2009dm} unfolding methods have been tested at the second step of the procedure.
These methods differ by the correction strategy, the way they rely on the shape of the MC spectrum at the particle level~(the SVD and IDS methods rely much less on this shape 
compared to the bin-by-bin method), as well as by their respective regularisation methods~(a local, significance-based regularization is used in IDS, while a regularisation based on
a singular value decomposition plus a constraint on the global curvature of the unfolded spectrum are used for SVD).

The potential bias of each of these unfolding methods has been studied using a data-driven closure test, relying on the shape comparison between data and MC at the reconstructed level.
For this test, a reweighting of the particle level MC spectrum~(used to build the transfer matrix) by a smooth function is performed, such that, after projection on the reconstructed MC axis,
a better agreement with data is observed.
The reweighted reconstructed MC is unfolded with each of the three methods, using the original transfer matrix~(without reweighting, like for the unfolding of the data) as input of the methods.
The comparison of the corresponding results with the reweighted particle level MC provides an estimation of the \emph{MC shape uncertainty} for each unfolding method.
The smallest uncertainty, at the $~1\%$ level, is obtained using the IDS method, while a larger uncertainty is obtained with SVD.
After a NLO/LO reweighting of the particle level MC shape in the input transfer matrix, the bin-by-bin method has an uncertainty similar to the one of IDS, while it was larger before.
The IDS method is used for performing the nominal correction of the data spectra.

The full set of uncertainties are propagated from the reconstructed to the unfolded level.
The statistical uncertainties are propagated using pseudo-experiments, where both the input data spectrum and the MC transfer matrix are statistically fluctuated and a covariance matrix is obtained.
Each component of the systematic uncertainty is propagated by performing shifts of the reconstructed spectrum by one standard deviation in the positive and negative direction respectively, redoing
the unfolding and comparing the results with the nominal unfolded spectrum.
The resolution uncertainty is propagated by performing a smearing of the nominal transfer matrix by the resolution uncertainty, re-doing the unfolding and comparing with the nominal result.

\section{Theoretical predictions and comparison with the data}

The unfolded experimental cross sections are compared with the NLO QCD prediction, corrected for non-perturbative effects.
Both the NLOJET++~\cite{Nagy:2003tz} and the POWHEG~\cite{Alioli:2010qp,Nason:2007vt}~(with parton shower switched off) generators were used for the hard scattering.
The CT10~\cite{Lai:2010vv} NLO parton distribution functions and the same values for the renormalization and factorization scales~(the transverse momentum of the leading jet) were used for
both programs.
An agreement between the results of the two programs at the few percent level was observed for the inclusive jet cross section, over all the rapidity regions.
They are also consistent for dijet events with both jets in the central region.
However, large differences are observed for dijet events with large rapidity separations between the two leading jets, for which the NLOJET++ prediction is unstable.
This prediction is much more stable when using a scale taking into account the rapidity difference between the two leading jets.
The results were also compared with predictions obtained using alternative PDFs: MSTW 2008~\cite{Martin:2009iq}, NNPDF 2.1~\cite{Ball:2010de} and HERAPDF 1.5~\cite{HERAPDF15}.

The non-perturbative correction factors are derived using PYTHIA~\cite{Sjostrand:2006za}, while the uncertainties are obtained from comparisons between different PYTHIA tunes,
as well as from the comparison with HERWIG++~\cite{Bahr:2008pv}.
These correction factors have been applied to the NLO predictions, in view of the comparison with the experimental cross sections.

\begin{figure}[t]
  \centering
  \includegraphics[width=0.592\textwidth]{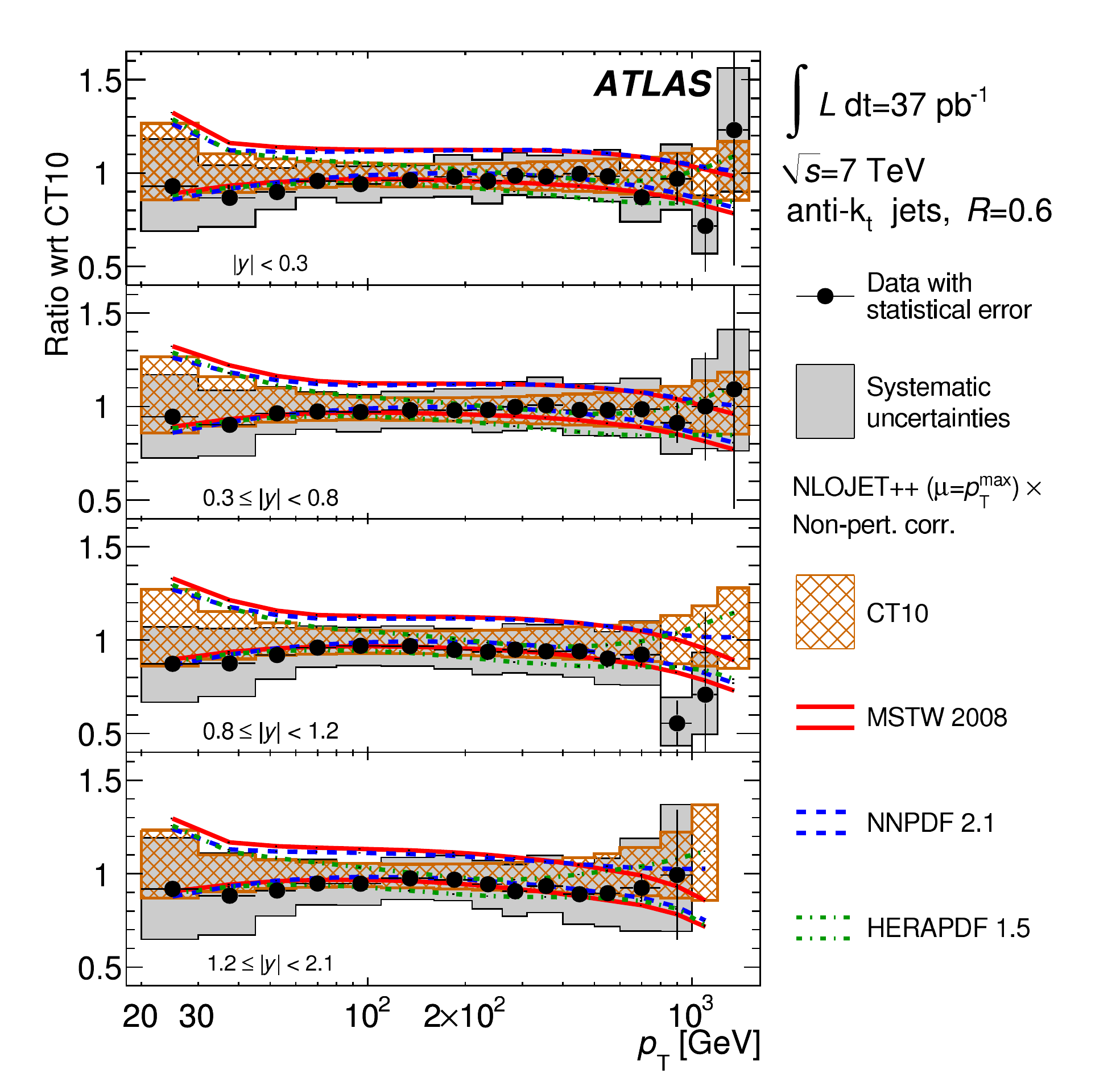}
  \hspace{-2.9cm}
  \includegraphics[width=0.592\textwidth]{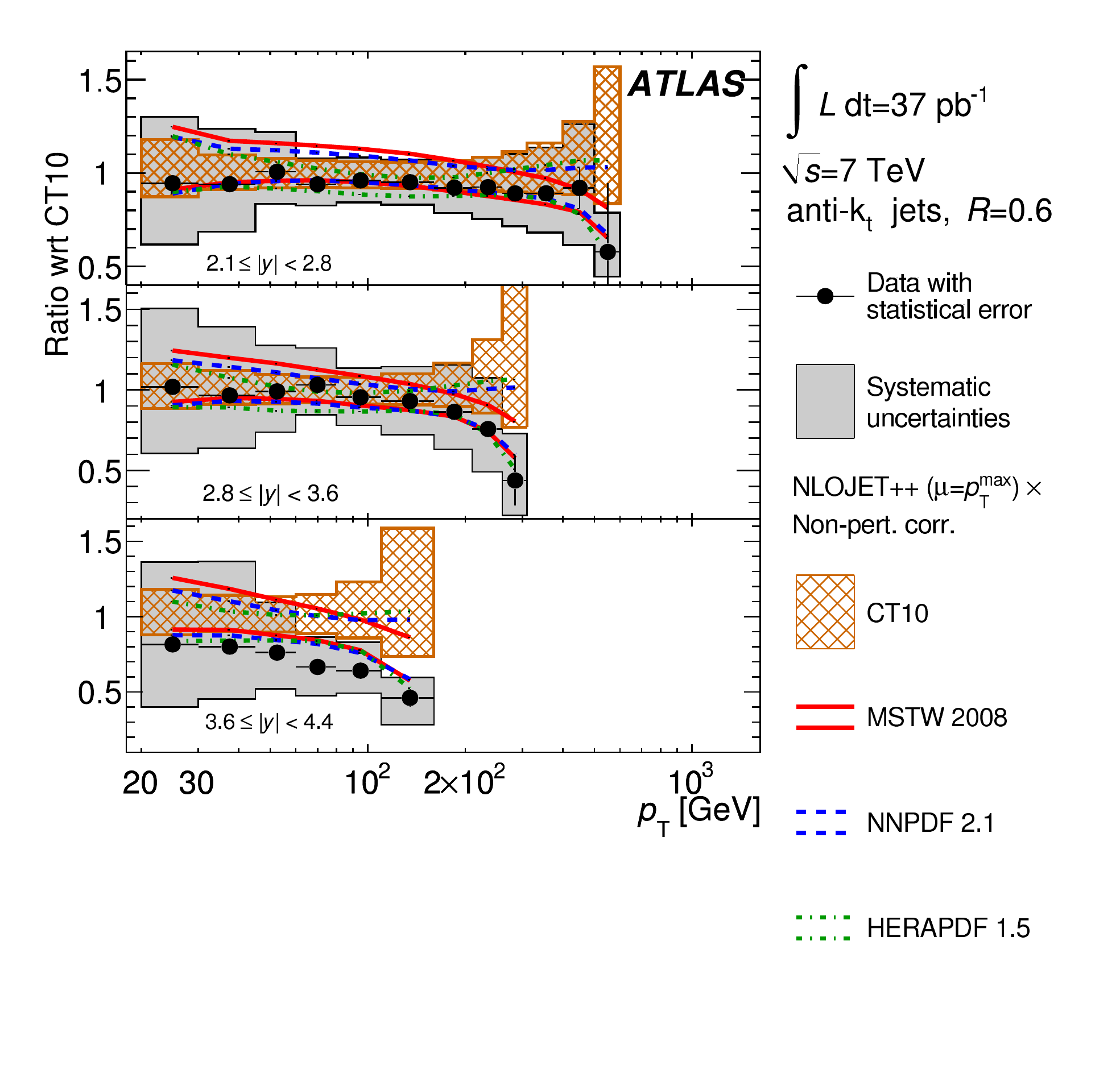}
  \caption{Relative comparison between the experimental inclusive jet cross section~\cite{Aad:2011fc} and the theoretical predictions obtained using NLOJET++ with various PDF sets.}
  \label{Fig:Comparison}
\end{figure}

Figure~\ref{Fig:Comparison} shows the relative comparison between the 2010 inclusive jet cross section, for jets with R=0.6, and the theoretical predictions obtained using NLOJET++,
with non-perturbative corrections.
Within the quoted uncertainties, the general agreement is good, except in the high transverse momentum region where the data start to be sensitive to the various PDF sets.

Comparisons have also been performed with the POWHEG prediction, showered with PYTHIA and respectively HERWIG.
Important differences between these two are observed, both for the inclusive and dijet cross sections.


\section{Conclusions}

The ATLAS experiment has performed several measurements of the inclusive jet and dijet cross sections, using data taken during 2010 as well as 2011.
These measurements, provided with the full information on uncertainties and correlations, allow for tests of QCD in phase space regions that were not covered previously.

{\raggedright
\begin{footnotesize}



\end{footnotesize}
}


\end{document}